\documentclass{webofc}
\usepackage[varg]{txfonts}   

\usepackage{subcaption}
\usepackage{xcolor}

\begin{document}
\title{A comparison of spectral reconstruction methods applied to non-zero temperature NRQCD meson correlation functions\footnote{Talk presented at {\em 
A Virtual Tribute to Quark Confinement and the Hadron Spectrum}, August 2-6 2021, University of Stavanger, Stavanger, Norway.}}

%
%

\author{\firstname{Thomas} \lastname{Spriggs}\inst{1}\fnsep\thanks{\email{996870@swansea.ac.uk}} \and
        \firstname{Gert} \lastname{Aarts}\inst{1,2} \and
        \firstname{Chris} \lastname{Allton}\inst{1} \and
        \firstname{Timothy} \lastname{Burns}\inst{1} \and
        \firstname{Rachel} \lastname{Horohan D'Arcy}\inst{3} \and
                \firstname{Benjamin} \lastname{J\"ager}\inst{4} \and
        \firstname{Seyong} \lastname{Kim}\inst{5} \and
        \firstname{Maria-Paola} \lastname{Lombardo}\inst{6} \and
        \firstname{Sam} \lastname{Offler}\inst{1} \and
        \firstname{Ben} \lastname{Page}\inst{1} \and
        \firstname{Sinead} \lastname{M. Ryan}\inst{7} \and
        \firstname{Jon-Ivar} \lastname{Skullerud}\inst{3} 
}
\institute{Department of Physics, Swansea University, Swansea SA2 8PP, United Kingdom
\and
           European Centre for Theoretical Studies in Nuclear Physics and Related Areas (ECT*) \& Fondazione Bruno Kessler Strada delle Tabarelle 286, 38123 Villazzano (TN), Italy 
\and
           Department of Theoretical Physics, National University of Ireland Maynooth, County Kildare, Ireland
\and
           CP3-Origins \& Danish IAS, Department of Mathematics and Computer Science, University of Southern Denmark, 5230 Odense M, Denmark
\and
           Department of Physics, Sejong University, Seoul 143-747, Korea 
\and
           INFN, Sezione di Firenze, 50019 Sesto Fiorentino (FI), Italy
\and
           School of Mathematics, Trinity College, Dublin, Ireland 
           }

\abstract{
We present results from the {\sc fastsum} collaboration's programme to determine the spectrum of the bottomonium system as a function of temperature. Three different methods of extracting spectral information are discussed: a Maximum Likelihood approach using a Gaussian spectral function for the ground state, the Backus Gilbert method, and the Kernel Ridge Regression machine learning procedure. We employ the {\sc fastsum} anisotropic lattices with 2+1 dynamical quark flavours, with temperatures ranging from 47 to 375 MeV.
  }
\maketitle
%


\section{Introduction}
\label{intro}

There has been a great deal of interest in onia systems in the context of heavy-ion collision experiments,
particularly since the proposal they behave as a proxy for the temperature \cite{Matsui:1986dk}. 
While initial work concentrated on the charmonium system,
interest turned to bottomonium for several reasons: these mesons are produced copiously in LHC heavy-ion collision 
experiments, they act as probes of the quark-gluon plasma, and results from the CMS Collaboration
indicate sequential suppression in this system \cite{CMS:2011all}.

The {\sc fastsum} collaboration has had a long programme of studying the bottomonium spectrum at non-zero temperature using the NRQCD method \cite{Aarts:2014cda},
principally using the Maximum Entropy Method \cite{Asakawa:2000tr}. We have determined both S- and P-wave masses
and determined upper bounds for the state's widths.
We extend this work here to include three new
analysis techniques to determine the bottomonium spectrum, including the widths of the states.
These approaches are: a Maximum Likelihood approach using a Gaussian spectral function for the ground state (see sec.~\ref{GML}),
the Backus Gilbert method (sec.~\ref{BG}), and the Kernel Ridge Regression machine learning procedure (sec.~\ref{KR}).
This contribution contains a progress report on this project, in which the long-term aim is to test and compare these and other methods,
using the same lattice dataset in order to extract the best estimates of the bottomonium spectrum at finite temperature.


\section{Lattice Method}
\label{lattice-method}

Because the $b$-quark's mass is larger than any other mass scale,
it can be approximated as a non-relativistic particle
and the NRQCD effective theory can be used for its dynamics. Note that this effective theory also holds in the case
of thermal QCD since the temperature reached in heavy-ion collisions is less than the $b$-quark mass.
We use an ${\cal O}(v^4)$ lattice implementation of NRQCD where $v = |{\bf p}|/M$ is the velocity of the
$b$-quark in the bottomonium's rest frame.
Details of the action and the tuning of the $b$-quark parameters can be found in \cite{Aarts:2014cda}.

NRQCD has simpler time evolution properties compared to relativistic theories,
it is an ``initial value'' problem with the quark propagating forward in time only.
The missing backward movers lead to a simpler spectral decomposition.
The meson correlator, $G(\tau)$, can be expressed in terms of the spectral function, $\rho(\omega)$, via
\begin{equation}
    G(\tau)  = \int^{\omega_{max}}_{\omega_{min}} \frac{ \mathrm{d}\omega}{2\pi} K(\tau, \omega) \rho(\omega),
    \qquad\qquad
    K(\tau,\omega)= e^{-\omega\tau}.
    \label{eqn:spect_relation}
\end{equation}
In NRQCD the kernel $K(\tau,\omega)$ takes the simple exponential form,
whereas in the relativistic case it is $\cosh[\omega(\tau-1/2T)]/\sinh(\omega/2T)$.

The challenge is to reconstruct $\rho(\omega)$ from $G(\tau)$. 
Eq.~(\ref{eqn:spect_relation}) illustrates the ``ill-posed'' nature of this problem:
$G(\tau)$ is typically known at ${\cal O}(10-100)$ data points,
whereas $\rho(\omega)$ is a continuous function which requires ${\cal O}(1000)$ points to correctly represent it.
This is a well-known problem and many techniques have been developed across several research fields to solve this problem. In the lattice context, it is important to compare and test these methods to ascertain which one(s) are most applicable.

{\sc fastsum}'s ``Generation 2L'' anisotropic lattice ensembles are used,
with a physical size of (32$a_s)^3\times(N_\tau a_\tau)$ where $a_\tau^{-1} = 5.997(34)$ GeV and $a_s/a_\tau=3.453(6)$ \cite{Edwards:2008ja,HadronSpectrum:2008xlg,Cheung:2016bym}.
The simulation is performed with 2+1 dynamical Wilson-clover quark flavours where the pion mass is $M_\pi = 236(2)$ MeV.
There are ${\cal O}(1000)$ configurations for all temperatures and the pseudocritical temperature is $162(1)$ MeV,
as measured from the inflection point of the renormalised chiral condensate \cite{Aarts:2020vyb}.
The temperatures and corresponding $N_\tau$ values are shows in Table \ref{tab:temperatures}.

\begin{table}[h]
 \centering
    \begin{tabular}{c||c|c|c|c|c|c|c|c|c|c|c}
    $N_\tau$ & 16 & 20 & 24 & 28 & 32 & 36 & 40 & 48 & 56 & 64 & 128 \\
    \hline
    T [MeV] & 375 & 300 & 250 & 214 & 187 & 167 & 150 & 125 & 107 & 94 & 47
    \end{tabular}
    \caption{Temporal lattice sizes and corresponding temperatures for the {\sc fastsum} Generation 2L ensembles \cite{Aarts:2020vyb}.
    The $N_\tau=128$ ensemble was kindly provided by the Hadron Spectrum Collaboration \cite{Edwards:2008ja,HadronSpectrum:2008xlg,Cheung:2016bym}.}
    \label{tab:temperatures}
\end{table}

\begin{figure}[h]
    \centering
    \includegraphics[width=0.9\textwidth]
    {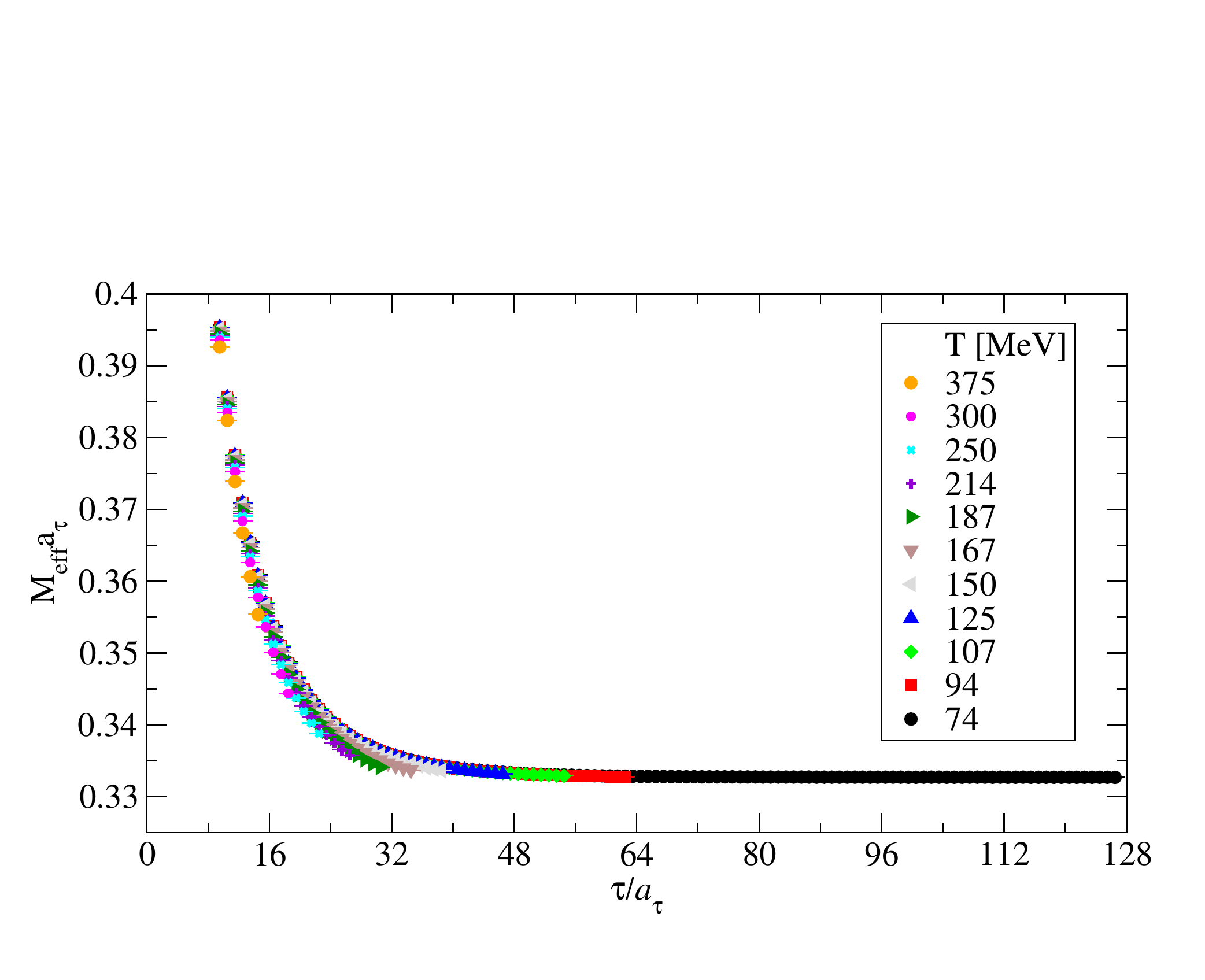}
    \caption{Effective mass plot in the $\Upsilon$ channel, for all temperatures considered. }
    \label{fig:eff-mass}
\end{figure}

To start the discussion we show in Fig.\ref{fig:eff-mass} the effective mass in lattice units for the $\Upsilon$ meson, defined via
\begin{equation}
M_{\text{eff}}(\tau) = \log\left(\frac{G(\tau)}{G(\tau+1)}\right),
\end{equation}
for all the temperatures considered. Some evidence 
of thermal effects can be observed.

\section{Gaussian Maximum Likelihood}
\label{GML}
In this approach the bottomonium spectral function is parametrised by a Gaussian ground state (see also \cite{Larsen:2019bwy}) and a single $\delta$-function to account for all spectral weight beyond the ground state,
\begin{equation}
    \rho_{\text{ansatz}}(\omega) = A_{\text{ground}} \exp\left(\frac{(\omega - M_{\text{ground}})^2}{2\sigma^2}\right) + A_{\text{excited}}\; \delta(\omega - M_{\text{excited}}).
\label{eqn:gaussian}
\end{equation}
The Gaussian ansatz for the ground state is chosen to allow for a finite width, which is expected in a thermal medium. The spectral weight beyond the ground state could be modelled with a function more sophisticated than a $\delta$-function, but the extra parameters in any such function would reduce the method's predictability.
Inserting Eq.\ (\ref{eqn:gaussian}) into Eq.\ (\ref{eqn:spect_relation}) leads to a closed form expression for the correlation function $G_{\text{ansatz}}(\tau)$.
We then apply the standard Maximum Likelihood method to determine the best fit parameters $\{ M_{\text{ground/excited}}, A_{\text{ground/excited}}, \sigma\}$.

\begin{figure}
    \centering
    \includegraphics[width=0.48\textwidth]{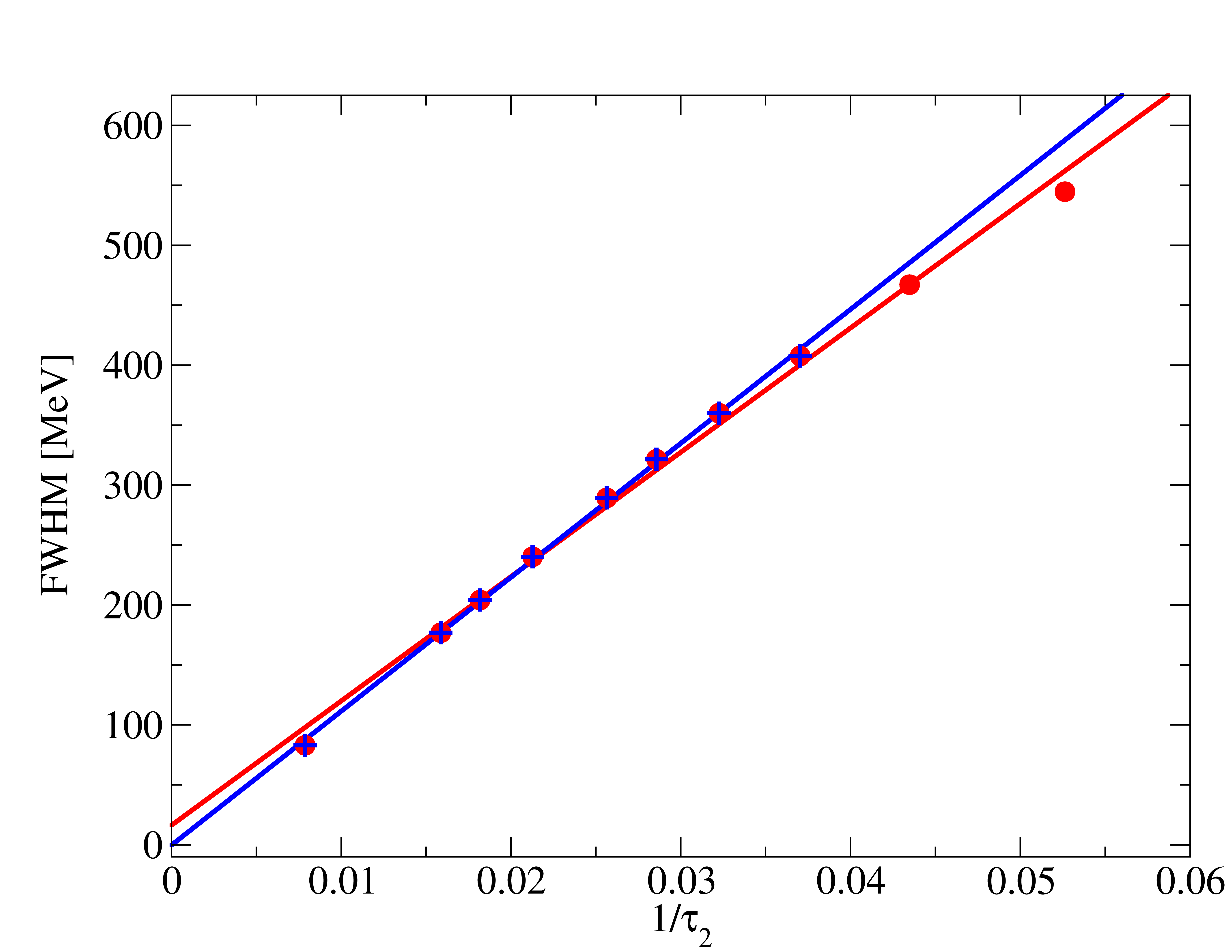}
    \includegraphics[width=0.48\textwidth]{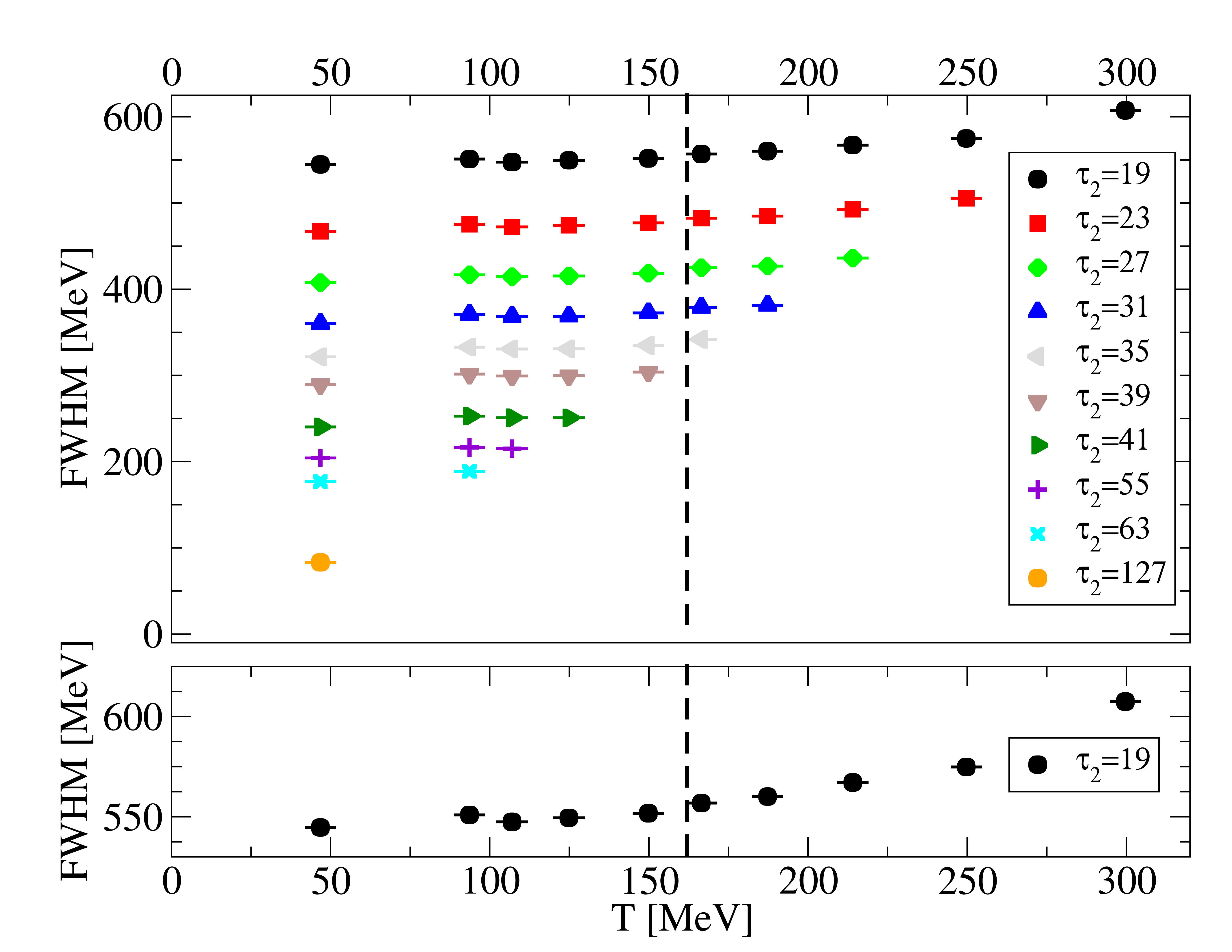}
    \caption{Left: Full width at half maximum (FWHM)
    for the $\Upsilon$ at $T=47$ MeV as a function of the inverse time window parameter $1/\tau_2$,
    with $\tau_1=8$ fixed. 
    The red (blue) line is a linear extrapolation on all (eight leftmost) data points.
    Right: FWHM of the $\Upsilon$ state for all temperatures for various $\tau_2$ values.
    The pseudo-critical temperature, $T_{\rm pc}=162(1)$ MeV, is shown by vertical dashed line.
    The lower pane contains a close-up using $\tau_1=8$ and $\tau_2=19$ for all temperatures.}
    \label{fig:gauss}
\end{figure}

In the Maximum Likelihood method, we need to chose a time window $\tau \in [\tau_1,\tau_2]$ to perform the fits.
This introduces a systematic effect which we study in Fig.~\ref{fig:gauss} (left).
Here the full width at half maximum (FWHM) is plotted against $1/\tau_2$ for the 
$\Upsilon$ channel at $T=47$ MeV with $\tau_1=8$  throughout. Note that we use temporal lattice units here, $a_\tau\equiv 1$, such that the $\tau$'s are integers.
We expect the best estimate of the width to be given as $1/\tau_2 \rightarrow 0$, since this isolates the ground state. However, at non-zero temperature the temporal extent of the lattice is naturally restricted. 
As can be seen, the width heavily depends on the time window. We consider two extrapolations
$1/\tau_2 \rightarrow 0$; the blue line 
is a linear fit to the eight leftmost (blue) data points, whereas the red line is a linear fit to all (blue and red) points. Taking these extrapolations at face-value yields a width estimate of $\sim 10$ MeV.

The effect of varying the time window $\tau \in [8,\tau_2]$ is shown in Fig.~\ref{fig:gauss} (right) for all temperatures. 
Note that results from the same time window are plotted with the same colour.
The variation in the width as the fit parameter $\tau_2$ is varied is substantial.
Further details can be found elsewhere \cite{tom_lat21}, including a more detailed analysis of the systematic effects. 
As a preliminary result we note from the right lower pane the  indication that the width increases above $T_{\rm pc}$
when results from the same time window but at different temperatures are compared.

\section{Backus Gilbert}
\label{BG}

The Backus-Gilbert method was introduced originally to solve an ill-posed problem relevant to geology \cite{Backus1968}.
It estimates a solution to Eq.~(\ref{eqn:spect_relation}), $\hat{\rho}(\omega)$, in a point-wise manner by sampling the target spectrum, $\rho(\omega)$, using a basis of resolution functions, $A(\omega, \omega_0)$, that are peaked around some $\omega_0 \in [\omega_{min}, \omega_{max}]$,
\begin{equation}
    \hat{\rho}(\omega_0) = \int_{\omega_{min}}^{\omega_{max}}A(\omega, \omega_0) \rho(\omega) d\omega.
    \label{eq:rho-hat}
\end{equation}
Ideally, these resolution functions are as close to the delta function $\delta(\omega-\omega_0)$
as possible.
They are expressed as a linear combination of the kernel function,
\begin{equation}
    A(\omega, \omega_0) = \sum_\tau c_\tau(\omega_0)K(\omega, \tau),
    \label{eq:basis-fns}
\end{equation}
where the coefficients, $c_\tau(\omega_0)$ are to be determined.
This means that, by combining Eqs.~\eqref{eq:rho-hat} and \eqref{eq:basis-fns}, the solution estimate
$\hat{\rho}(\omega_0)$ is obtained from a linear combination of the original correlation functions, $G(\tau)$,
\begin{equation}
    \hat{\rho}(\omega_0) = \sum_\tau c_\tau(\omega_0)G(\tau).
\end{equation}
The coefficients $c_\tau(\omega_0)$ can be determined by the Dirichlet least-squares criterion of minimising the distance
between the resolution functions and the delta function,
\begin{equation}
    J(\omega_0) = \int_{\omega_{min}}^{\omega_{max}} [A(\omega, \omega_0) - \delta(\omega - \omega_0)]^2 d\omega,
\end{equation}
leading to 
\begin{equation}
    \mathcal{K}_{\tau \tau'} \cdot c_{\tau'}(\omega_0) = K(\omega_0, \tau) \hspace{1cm} \text{where}   \hspace{2mm} \mathcal{K}_{\tau \tau'} = \int_{\omega_{min}}^{\omega_{max}} K(\omega, \tau) K(\omega, \tau')d\omega.
\end{equation}
The variance in the solution is also a function of $c_\tau(\omega_0)$ and is given by
\begin{equation}
    \text{Var}[\hat{\rho}(\omega_0)] = \sum_{\tau, \tau'} c_\tau \Sigma_{\tau,\tau'}c_{\tau'},
\end{equation}
where $\Sigma_{\tau,\tau'}$ is the covariance in $G(\tau)$.

\begin{figure}[t]
    \centering
    \includegraphics[width=0.48\textwidth]{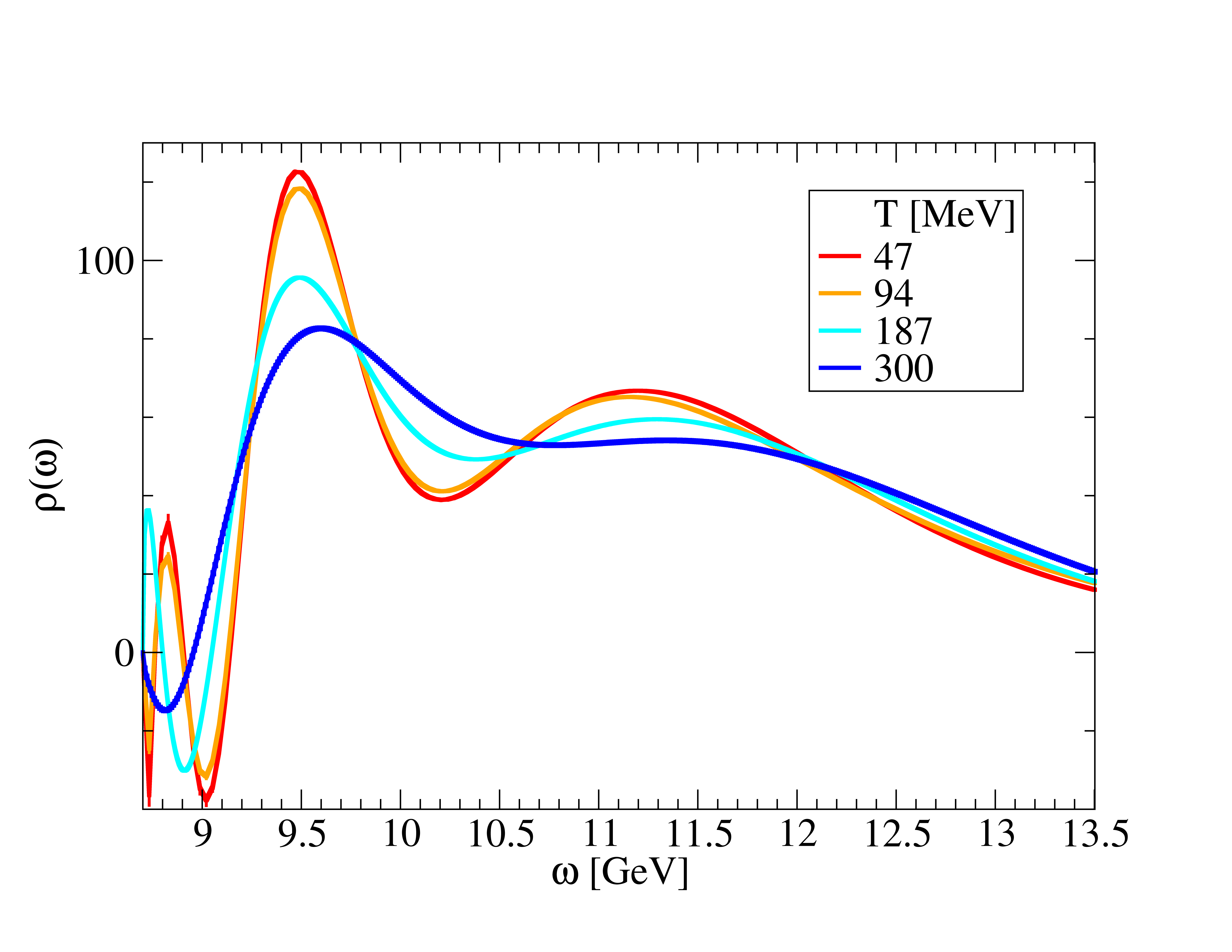}
    \includegraphics[width=0.48\textwidth]{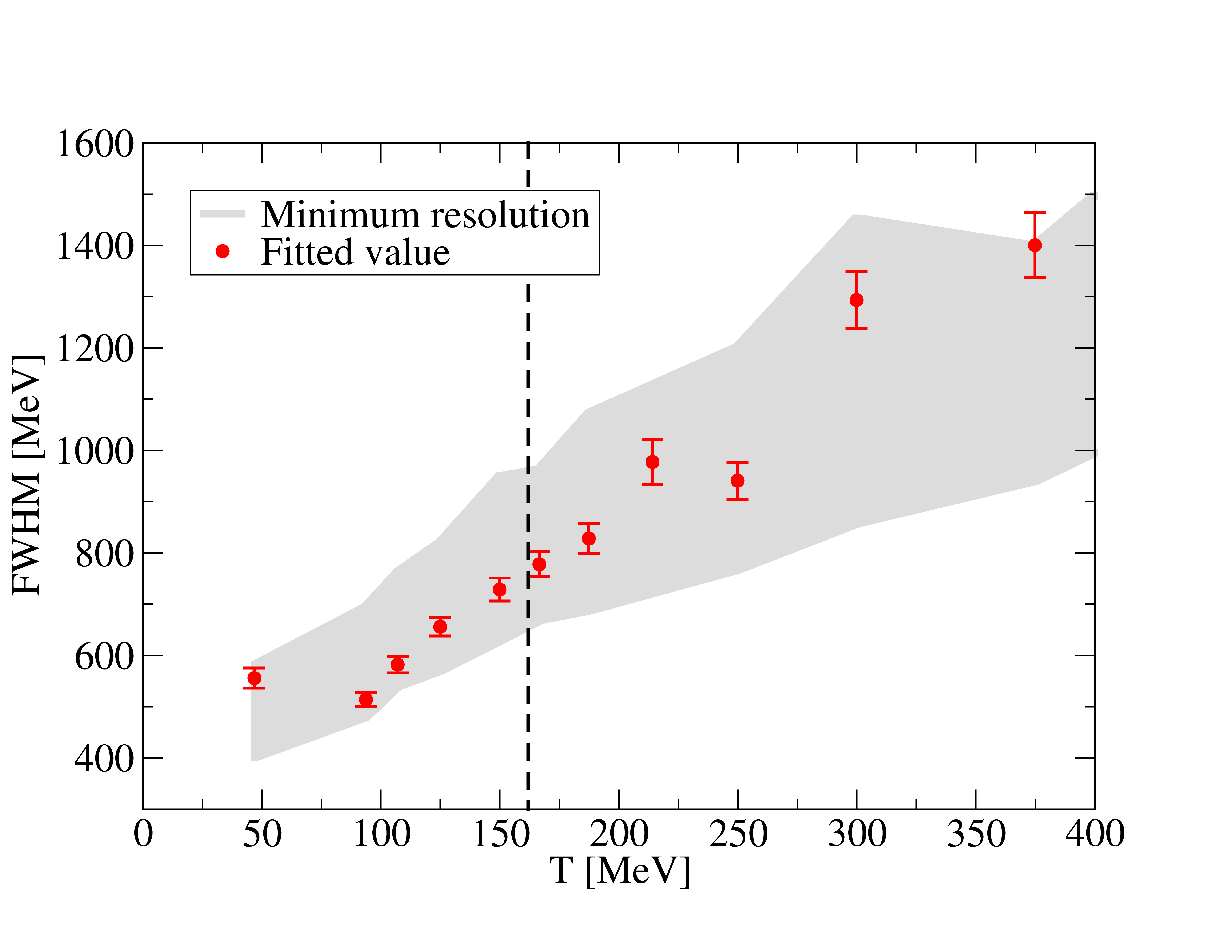}
    \caption{Left: Reconstructed spectral functions of $\Upsilon$ from the Backus-Gilbert method for a sample of temperatures.
    Right: The ground state's FWHM against temperature with the vertical dashed line indicating $T_{\rm pc}$.
    The grey band shows the minimum resolution, and its error band, that the resolution functions could achieve at a given temperature.}
    \label{fig:BG}
\end{figure}

Reconstructed spectral functions for a sample of temperatures are shown in Fig.~\ref{fig:BG} (left).
From this we can see little change in the spectrum between 47 and 97 MeV, but then an increase in mass
and a large apparent increase in width at larger temperatures.
We note that the features below 9 GeV are artefacts of the method.
In Fig.~\ref{fig:BG} (right) the FWHM is plotted against temperature where the error bars are statistical only.
At first sight, there appears a clear increase in the FWHM with temperature.
However, also plotted in the grey band is the FWHM of the resolution function $A(\omega,\omega_0)$
and its error bound calculated at $\omega_0 = M_{\text{ground}}$.
This represents the lower bound in the width that the method can resolve
-- the method is incapable of predicting widths below this bound, and
only widths which lie above this bound can be taken as predictions.
Since the bottomonium widths calculated using this method follow this lower bound,
the Backus Gilbert method is currently incapable of resolving the bottomonium width.
Further discussion can be found in \cite{ben_lat21,inpreparation}.


\section{Kernel Ridge Regression}
\label{KR}

As is common in machine learning paradigms, kernel ridge regression (KRR) infers a prediction of a quantity based on previously observed training data. In the case of solving the inversion problem for NRQCD, mock data with known spectral functions are generated as training data.

Details of the KRR approach in the context of NRQCD can be found in \cite{sam-old}.  Here we note that we modified our method how 
the input data, the set of correlators $G_i(\tau)$ (where $i=1,2,\ldots, N_{\rm train})$, is combined into the matrix (or kernel) $\mathbf{C}$ before the regression procedure. In contrast to \cite{sam-old}, we use here the following encoding for the matrix elements of $\mathbf{C}$, 
\begin{equation}
C_{ij} = \exp \left( -\gamma \sum_n \left[\frac{G_i(\tau_n) - G_j(\tau_n)}{\overline{G}(\tau_n)}\right]^2\right),
\end{equation}
where the sum $n$ runs over all time slices (starting at $n=4$), 
$\gamma$ is a hyperparameter that sets the correlation length in the space of correlators,
and $\overline{G}(\tau_n) = 1/N_{\rm train} \sum_i G_i(\tau_n)$ is the average correlator of the training data at time $\tau_n$.

\begin{figure}[t]
    \centering
    \includegraphics[width=0.48\textwidth]{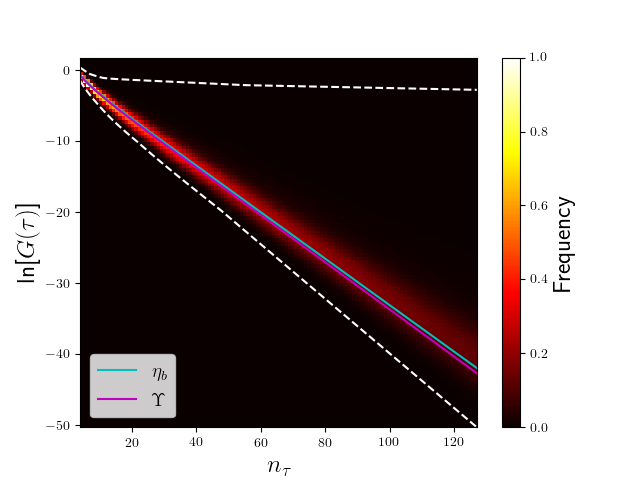}
    \includegraphics[width=0.48\textwidth]{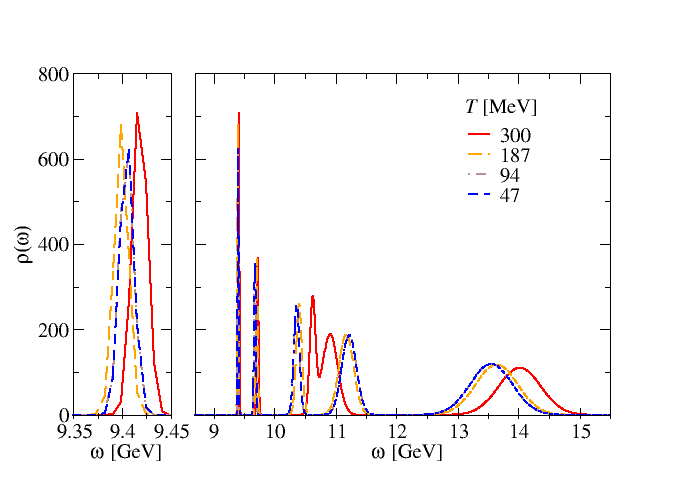}
    \caption{Left: Frequency of central values generated to form the set of mock correlation functions overlaid with the actual correlation functions, the data is constrained to lie within the dashed lines and be most frequent near the true data. 
    Right: Reconstructed spectral functions at a sample of temperatures with a zoom in of the ground state peak highlighted to observe the temperature dependence.}
    \label{fig:KRR}
\end{figure}

We continue with some remarks on the generation of mock data, which has applications beyond KRR.
KRR works by inferring a spectral function from a training set of mock spectral functions weighted by how much the concomitant correlation function resembles the true lattice correlation function. Clearly then, the closer the training correlation functions are to the actual lattice correlation functions obtained via numerical simulations, the more relevant training spectral functions the KRR has to draw upon to infer a spectral function. Building on \cite{sam-old}, work has been done to increase the overlap of the training data and the lattice data.
The overlap between the new training data and the actual lattice correlators is shown in Fig.~\ref{fig:KRR} (left).
Here, the solid lines are the lattice correlators obtained in the simulations at a given temperature ($N_\tau=128$ in this case). The heatmap shows the distribution of the training data around these values and the dashed lines enclose the region inside which all training data lies. As can be seen, the mock data tracks the actual correlators well. 
More details can be found in \cite{sam_lat21}. 

Resulting spectral functions are shown in Fig.~\ref{fig:KRR} (right). We observe ground state peaks at all four temperatures shown, with some temperature dependence visible in the left pane. It is noted that the mass of both the ground state and first excited state lie below the experimental values. Further discussion is presented in  \cite{sam_lat21,inpreparation}.


\section{Conclusion}
\label{conclusion}

This work presents a progress report of the {\sc fastsum} collaboration's programme to apply a number of methods
to reconstruct the bottomonium spectral functions from NRQCD lattice correlation functions.
The three methods used are
a Maximum Likelihood approach using a Gaussian spectral function for the ground state;
the Backus Gilbert method; and the Kernel Ridge Regression machine learning procedure.
The results presented are preliminary and indicate the substantial systematic uncertainties.

Further work to analyse these systematic effects is in progress \cite{inpreparation}.


\section*{Acknowledgements}

This work is supported by STFC grant ST/T000813/1.
RHD has been supported by a Maynooth University SPUR scholarship.
SK is supported by the National Research Foundation of Korea under grant NRF-2021R1A2C1092701 funded by the Korean government (MEST).
BP has been supported by a Swansea University Research Excellence Scholarship (SURES).
This work used the DiRAC Extreme Scaling service at the University of Edinburgh, operated by the Edinburgh Parallel Computing Centre on behalf of the STFC DiRAC HPC Facility (www.dirac.ac.uk). This equipment was funded by BEIS capital funding via STFC capital grant ST/R00238X/1 and STFC DiRAC Operations grant ST/R001006/1. DiRAC is part of the National e-Infrastructure.
This work was performed using PRACE resources at Cineca via grants 2015133079 and 2018194714.
We acknowledge the support of the Supercomputing Wales project, which is part-funded by the European Regional Development Fund (ERDF) via Welsh Government,
and the University of Southern Denmark for use of computing facilities.
We are grateful to the Hadron Spectrum Collaboration for the use of their zero temperature ensemble.

\end{document}